# A Proximity based Retransmission Scheme for Power Line Ad-hoc LAN

Chitta Ranjan Singha

Assam Engineering Institute, Guwahati, Assam
crsingha@yahoo.com

## ABSTRACT

*Power line as an alternative for data transmission is being explored, and also being used to a certain extent. But from the data transfer point of view, power line, as a channel is highly dynamic and hence not quite suitable. To covert the office or home wiring system to a Local Area Network (LAN), adaptive changes are to be made to the existing protocols. In this paper, a slotted transmission scheme is suggested, in which usable timeslots are found out by physically sensing the media. Common usable timeslots for the sender-receiver pair are used for communication. But these will not ensure safe packet delivery since packets may be corrupted on the way during propagation from sender to receiver. Therefore, we also suggest a proximity based retransmission scheme where each machine in the LAN, buffers good packet and machines close to the receiver retransmit on receiving a NACK.*

## 1. INTRODUCTION

The increasing demand for high-speed data service has motivated to use power line as alternative communication media. Low voltage power supply networks are available in almost all households all over the world. The electrical wiring system of office or home can be converted to a Local Area Network (LAN) besides connecting the electrical appliances. Using home/office-wiring system to set up a LAN give certain advantages. It uses the existing wiring as the communicating media and hence saves extra cost of wiring. Every power point can be used to plug in a machine.

A power line network has a number of similarities with the wireless network. Both are noisy and unpredictable and both allow certain degree of mobility. Wireless Mobile ad hoc network (MANET) [1] has become popular and are being widely used. In an ad-hoc network, machines join, work and leave at will. Turning the home/office wiring system to an ad hoc LAN will have the advantage of scalability. But ad hoc networks have their own challenges to overcome and using power line as the communicating media add to the problems.

The major problem of a power line media is noise. Noise in power line is both synchronous and asynchronous. Since noise is inherent in power line and cannot be minimized, avoiding noise will increase throughput. This can be done by dividing the power cycle into a number of timeslots and by transmitting during the timeslots where the effect of noise is minimum.

LANs are generally setup using reliable media like co-axial and UTP/STP cables. So it is assumed that a packet sent by a machine reaches its destination without suffering data loss. However this assumption cannot be accepted for a highly unreliable media like the power line cable. Though the commonly used Ethernet LAN uses CSMA/CD as the MAC protocol, for power line networks CSMA/CA is being used. The HomePlug Power line Alliance, Inc., has been marketing home networking products and is the leader in this field. However independent studies [2][3] have found that though HomePlug has been able to overcome the effect of signal distortion with OFDM, and minimize the effect of line synchronous periodic impulse noise by applying FEC, the asynchronous impulse noise of long duration, degrade the performance of HomePlug to a great extent. Our time division based slotted transmission scheme is to minimize the effect of synchronous periodic noise and the proximity based retransmission scheme is to overcome the effect of long duration asynchronous noise. In the next sections, we first discuss power line as communication media, secondly the model of a power line LAN and then we discuss our adaptive approach.

## 2. RELATED WORKS

In the recent times a lot of research works have been done to analyze the power line for measurement, characterization and simulation of noise, signal attenuation and reflection [4][5]. Power line has the inherent characteristics of cyclic variation of channel and noise with the phase of the AC line cycle. Kyong-Hoe Kim et al [6] put forward a dynamic channel adaptation scheme that exploits the cyclic variation of power line to achieve high throughput.

In the field of power line communication, HomePlug [7] has done most advanced works. HomePlug is an industrial organization that now comprises of 70 countries was formed in 2000 to develop and standardize specification for home networking technology using existing power line wiring. The first release of HomePlug, called HomePlug 1.0, supports raw data rate up to 14 Mbit/s. The next version called the HomePlug AV (HPAV) supports raw data rate up to 150 Mbit/s. In the PHY, it employs Orthogonal Frequency Division Multiplexing (OFDM) over a bandwidth from 1.8 to 30 MHz using 917 useful sub-carriers, Forward Error Correction (FEC), error detection, data interleaving and Automatic Repeat Request (ARQ).

Tonello et al [8], in their work, have analyzed the resource allocation problem in a power line communication system similar to the HomePlug AV specifications, in a multi user scenario. They have determined the optimal time slot duration for various amounts of overhead required by the PHY layer, by taking into account both the cyclo-stationary noise and the cyclic variation of the channel response.

The CSMA/CA protocol has been used for sensor networks and for other wireless communications. For power line communication also CSMA/CA is being used. Sung-Guk Yoon et al [9] have suggested a sub-carrier based CSMA/CA protocol in OFDM based PLC system to improve throughput. The proposed protocol divides the whole bandwidth into several sub-channels and each station joins its best sub-channel. For communication, each sub-channel performs CSMA/CA independently, which will reduce the collision probability.

## 3. POWER LINE AS COMMUNICATION MEDIA

Power line, which is designed and installed to supply 50/60Hz power waves to home/office, suffers from many disadvantages. The channel characteristic is volatile and unpredictable. The main disadvantages are noise, attenuation and signal distortion.

## 3.1 Noise

Power line noise is both time and frequency dependent [6]. Unlike the other telecommunications channels, the power line channel does not represent an Additive White Gaussian Noise (AWGN). The power line noise can be classified as follows [10]:

- **Background colour noise:** This is the common type of noise present in all types of communication channels. It has a comparatively low power spectral density (PSD).
- **Periodic synchronous impulse noise:** This noise is created due to devices like thyristor controlled light dimmers. It is synchronous with the AC supply frequency. These are impulses of several tens of volts of twice the AC frequency occurring at every half cycle.
- **Non-periodic asynchronous impulse noise:** This type of noise is due to various switching operations and contact switching on loads. It has high and short voltage peaks of up to 2kV and of 10 - 100μs in length.
- **Periodic asynchronous impulse noise:** This noise is due to switched power supply in PCs and electronics equipments. It contains impulses that are repeated between ca. 50 and 200 kHz. They are not synchronized with AC supply.

## 3.2 Attenuation

Attenuation of signals in a home/office power line may occur for various reasons. It is dependent on the type of cable used. Attenuation is more significant in longer lines and increases with increasing frequency [11]. The signal becomes weaker also while passing through the breaker panels.

## 3.3 Signal distortion

Because of the presence of a number of junction and branching points, the signal undergoes reflection. Due to different path lengths and reflection coefficients, signals arrive at the receiver staggered in time and with different amplitudes, and the receiver receives replicas of the original transmitted signal with different attenuations and time delays. This is called the multi-path distortion.

## 4. MODEL OF A POWER LINE LAN

Fig. 1 shows a probable model of a power line LAN. Four branches of various lengths are connected to the breaker point B. Each branch has multiple numbers of machines (1 - 12) connected along with electrical components (not shown) e.g. bulb, tube, TV etc. The important points that are to be noted here are,

>Point 1: All the machines are in the same collision domain.
>Point 2: Each machine may experience different channel characteristic than the others and the channel characteristics changes with time.
>Point 3: Due to high attenuation over the power line, the noise is location dependent [7n].

Point 1 implies that a message transmitted from a machine reaches all other machines, though there will be occasional packet corruption at certain points of the channel due noise. Point 2 implies that state of a packet received by a receiver will depend on the state of the channel at that instant which is also different for different machine. Point 3 implies that any instability at certain point of the channel, affects only the machines at its vicinity, all other machines remaining unaffected. State of the packet received depends on the path from the transmitter to receiver and not on the state of any other part of the network. So if machine 11 transmits, a noise burst at point X will make the packet received by machines 1, 2 and 3 corrupted, all other machine remaining unaffected by this. This phenomenon will be more prominent in big LANs covering a big building or the campus of an organization.

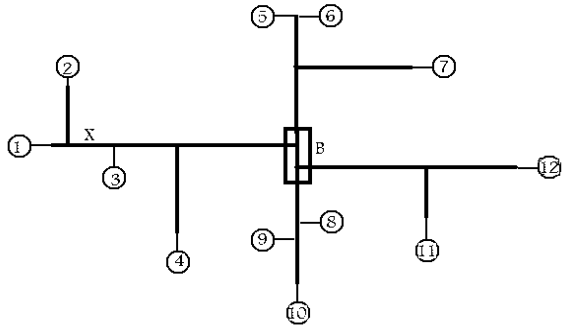

Figure 1. A Power line LAN

## 5. AN ADAPTIVE APPROACH

In the following, we discuss a slotted transmission to overcome synchronous periodic impulse noise and then the proximity based retransmission scheme.

### 5.1 Slotted transmission

We use CSMA /CA [12] as the MAC protocol, but modifications are suggested to suit the noisy channel like the power line.

- All the machines in the LAN are synchronized with power cycle.
- A power cycle is divided into 64 timeslots. A machine senses the carrier and is allowed to transmit at the beginning of a timeslot. Figure 2 shows the timeslots.
- In the PHY, during idle time, the machines sense the media for complete power line cycles (for 3 cycles) to detect noise and to find the usable timeslots. Information of the usable timeslots are stored and used to transmit and receive for the next 300 power cycles (6s). Since the channel conditions are different at different parts of the LAN, the list of usable timeslots will also be different for different machines.
- Considering a threshold value of noise power, a timeslot may be considered usable or unusable.
- Communication between two machines is done in common usable timeslots.

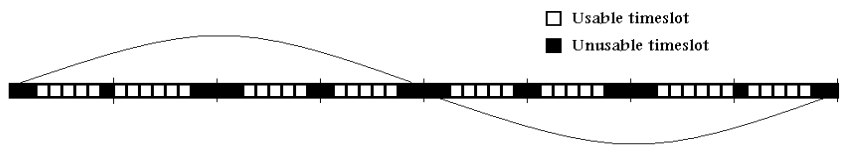

Figure 2. Timeslots of a Power Cycle

### 5.2 Proximity list

An asynchronous impulse noise of long duration, in the vicinity of the receiver and on the way from sender to receiver, will destroy the transmitted packet. Also the transmitted signal may become very weak and unrecognizable on the way from the sender to receiver in a wire joint

29

International Journal of Distributed and Parallel Systems (IJDPS) Vol.1, No.2, November 2010due to multi-path propagation. Due to volatile characteristics of the channel, the sender/receiver pair may not have a good number of usable timeslots for retransmission. If the transmitted packet is stored by a machine close to the receiver and retransmitted, probability of the packet reaching its destination will improve. This will also take less timeslots compared to a retransmission initiated and completed by the original sender. The number of common usable timeslots decides closeness of a pair machines here.

- The machines in the LAN, maintain a list of machines, which have most number of common usable timeslots. We call it the **proximity list**.
- During a communication processes, all the machines will observe the RTS sent by other machines and will update their proximity list. For a machine, at the beginning, the proximity list will be blank, but as communications go on it will be complete. The list may contain 2/3 entries.
- The machines will broadcast their usable timeslot information time to time.

### 5.3 Adaptive CSMA/CA

To start a communication, the sender senses the media and if found free, waits for starting of the next timeslot. Before transmitting a data packet, the sender machine sends a RTS packet. In the RTS packet, the sender passes the usable timeslot information to the receiver along with duration and other information. **On receiving the RTS, the other machines in the LAN defer and wait for the CTS**. The receiver calculates the required timeslots for transmission including ACK/NACK, from common timeslots of the sender and receiver, *Duration* field of the RTS and its existing NAV setting. The required timeslot include two additional timeslots (the reason is discussed on the later part). If a suitable window is found the CTS is sent in the same timeslot. The CTS include the required timeslots and *Duration* fields. By observing the CTS, other machines will set their NAV and defer. This way **timeslots are reserved for the communication** as shown in Figure 3. The sender will transmit in the reserved timeslots. The timeslots in between the reserved slots will be free for contention since common usable timeslots may be different for different sender receiver pair.

As the Sender transmits, all machines will receive the packet and verify the CRC. If found correct, the packet will be adopted and stored temporarily for retransmission, otherwise it will be dropped.

The receiver, after receiving a data packet, will verify the CRC. If no error is detected, the ACK packet is sent otherwise a NACK is sent. The ACK packet contains the CRC of the packet received, sender's address, receiver's address and the CRC. The NACK packet contains the CRC of the packet received, sender's address, top two addresses form its proximity list (PA1 & PA2), the receiver's timeslot information, Duration and the CRC (Figure 4).

On receiving the ACK for the packet, the machines, which had adopted the packet, will drop the packet. On receiving a NACK, all the machines whose addresses are not in the NACK, will drop the adopted packet.

The NAV will include two additional timeslots for the machines with addresses in the NACK to respond. The machines in the NACK will respond only when they have the valid packet buffered in them. The machine with PA1 in the NACK will respond in the first timeslot and if it fails to respond then only the machine with PA2 respond in the second timeslot.

30

International Journal of Distributed and Parallel Systems (IJDPS) Vol.1, No.2, November 2010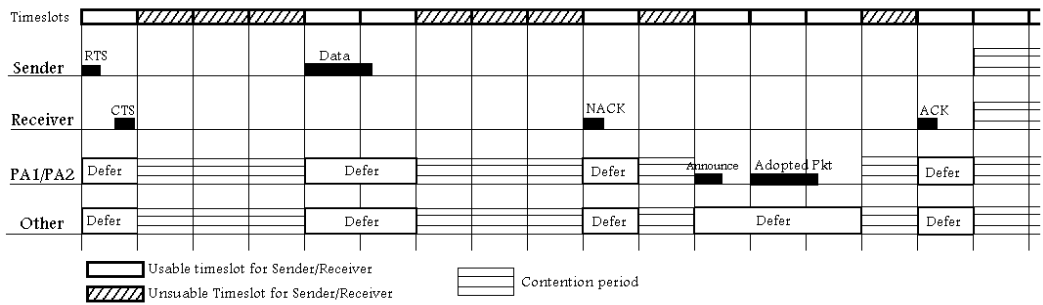

Figure 3. Slotted Packet Transmission

PA1 (or PA2) machines will respond by **announcing** the required additional timeslots for retransmission, which it will calculate from the receiver's timeslot information and Duration field in the NACK packet and its own timeslot information. The retransmission will be carried out during reserved additional timeslots while other machines will set the extended NAV and defer. On receiving a retransmitted packet, the receiver will send an ACK to the original sender, but no NACK for a retransmitted packet. If the original sender receives no ACK for the retransmission, it goes for contention to retransmit.

Any duplicate transmission will be handled by CSMA/CA as usual.

The functions performed by different machines in the LAN can be given as follows:

**Sender:**

Send_RTS
Wait_for_CTS     *'same timeslot'*
If (CTS_Received) {
   Transmit_data_packet
   Wait_for_ACK/NACK   *'NAV period'*
   If (ACK_received) {End}
   Else If (NACK_received) {
      Wait_for_ACK      *'extended NAV period'*
      If (ACK_received) {End}
      Else {Contention, Redo}}
   Else {Contention, Redo}}
Else {Contention, Redo}

**Receiver:**

If (Receive_RTS) {
  Send_CTS    *'same timeslot'*
  Wait_for_data_packet
  If (Data_packet_received) {
    Check_CRC_for_error
    If (No_error) {Send_ACK, End}
    Else {Send_NACK
      Wait_for_Announce   *'Next two slots'*
    If (No_announce){End}
    Else {Wait_for_adopted_packet   *'Extended NAV'*
      If (Adopted_Packet_received){Check_CRC_for_error

31

```
                     If (No_error) {Send_ACK_to_original_sender, End}
                     Else {End}}
               Else {End}}
      Else{End}}}
```

**Others:**

```
If (RTS_Received){Wait_for_CTS    'Same timeslot'
   If (CTS_received){
      Defer_in_reserved_slots
      Wait_for_data_packet    'Next reserved slot'
      If (Data_packet_received){Check_CRC_for_error
         If (Error){Drop_packet, End}
         Else {Store_packet, Wait_for_ACK/NACK 'Next reserved slot'
            If (ACK_received){Drop_packet, End}
            Else if (NACK_received){Check_PA1/PA2_with_own_address
               If(Address_match){Announce, Retrnasmit, Drop_packet}
               Else {Drop packet, Wait_for_announce
                  If (Receive_announce){Defer_in_extended_NAV, End}
                  Else {End}}}
            Else {Drop_packet, End}}}
      Else {End}}}
Else{End}
```

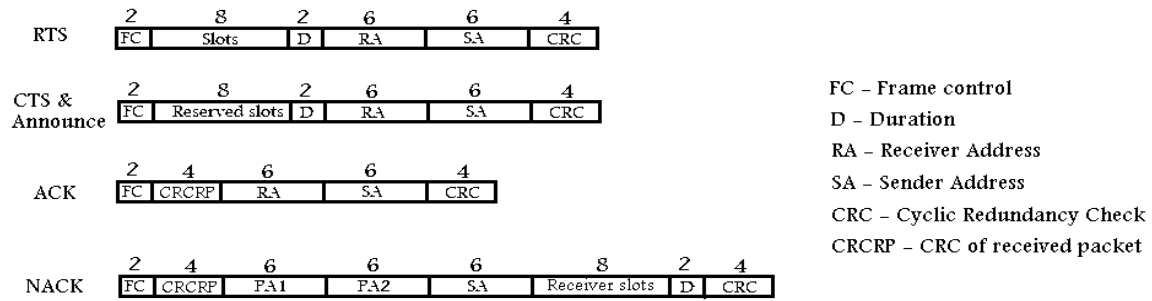

Figure 4.  Frames

# 6. CONCLUSION

The retransmission scheme will work for both small and big LANs but will be more effective for bigger LANs where signal will suffer more attenuation while propagating through a longer noisy channel. It will make the network immune to the periodic and non-periodic impulse noise to a large extent. Unless it is a continuous impulse noise like one created by a vacuum cleaner [13], large duration noise created by appliances like electric stove, electric iron etc will be handled efficiently. Avoiding noise by using usable timeslot will minimize packet corruption and hence will increase throughput. Also allowing multiple sender-receiver pair to communicate simultaneously will increase the channel usage. For reliable communication, proper synchronization of the machines is required.

Retransmission by the sender itself will not ensure packet delivery since the cause of packet corruption may still exist. So retransmission by a more reliable source will increase the rate of packet delivery. The problem may be the NAV setting to count multiple times. But a programmable counter may be designed for multi slot count setting.

## REFERENCES

[1] http://www.novaroam.com/Inside.asp?n=Technology&p=MANET

[2] Rohan Murty, Jitendra Padhye, Ranveer Chandra, Atanu Roy Chowdhury, and Matt Welsh, *Characterizing the End-to-End Performance of Indoor Power line Networks*, http://www.havard.edu/~rohan/papers/poweline-tr.pdf

[3] http://www.broadbandhomecentral.com/bbhl/homeplugconclusions.html

[4] T.Esmailian, F.R.Kschischang and P.G.Gulak, *Charateristics of in-building Power lines at High Frequencies and their Channel Capacity*, http:// www.isplc.org/docsearch/Proceedings/2000/pdf/0602_001.pdf

[5] Thomas Banwell, *A Novel Approach to the Modeling of the Indoor Power line Channel Part 1: Circuit Analysis and Companion Model*, IEEE Transaction on Power Delivery, VOL 20, No. 2, April 2005.

[6] Kyong-Hoe Kim, Han-Byul Lee, Yong-Hwa Kim, Seong-Cheol Kim, *Adaptation for Time-varying Power line Channel and Noise Synchronized with AC Cycle,* 978-1-4244-3790-0/09,2009 IEEE pp250- 254

[7] HomePlug Alliance. HomePlug_1.0 Technical Whitepaper_Final, http://www.homeplug.org/

[8] Tonello, A.M.; Cortes, J.A.; D'Alessandro, S. *Optimal time slot design in an OFDM-TDMA system over power line time variant* channels Power Line Communications and Its Applications, 2009. ISPLC 2009. IEEE International Symposium on , vol., no., pp.41-46, March 29 2009-April 1 2009

[9] Sung-Guk Yoon; Daeho Kang; Saewoong Bahk; *OFDMA CSMA/CA protocol for power line communication,* Power Line Communications and Its Applications (ISPLC), 2010 IEEE International Symposium on , vol., no., pp.297-302, 28-31 March 2010

[10] Jiri Misurec, *Interference in data communication over narrow-band PLC,* IJCSNS International Journal of Computer Science and Network Security, VOL.8 No.11, November 2008.

[11] Takatoshi MAENOU, Masaaki KATAYAMA, *Study on Signal attenuation Characteristics in Power Line Communications,* 1-4244-0113-5/06, 2006 IEEE.

[12] Pablo Brenner, *A Technical Tutorial on IEEE 802.11 Protocol*, http://www.sss-mag.com/pdf/802_11tut.pdf

[13] Gen Marubayashi, *Noise Measurement of Residential Power line*, http:// www.isplc.org/docsearch/Proceedings/1997/pdf/0551_001.pdf

## Author:

Chitta Ranjan Singha, Lecturer (Selection Grade) in the Department of Electronics & Telecommunications, in Assam Engineering Institute. He graduated from the Visvesvarya Regional College of Engineering (now VNIT), Nagpur and completed his M.Tech. from Tezpur University. His research area is Power Line Communication.

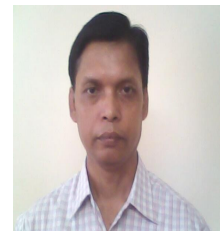